\documentclass[%
preprint,
amsmath,amssymb,
aps,
pra,
]{revtex4-2}

\usepackage{graphicx}
\usepackage{amsmath}
\usepackage{multirow}
\usepackage{color}

\begin{document}

\author{Lorenz S. Cederbaum}
\affiliation{Theoretische Chemie, Physikalisch-Chemisches Institut, Universit\"at Heidelberg, Im Neuenheimer Feld 229, Heidelberg D-69120, Germany}
\email{Lorenz.Cederbaum@pci.uni-heidelberg.de}

\title{Cooperative molecular structure in polaritonic and dark states}

%\date{\today}
\newpage

\begin{abstract}
 
An ensemble of identical, intrinsically non-interacting molecules exposed to quantum light is discussed. Their interaction with the quantum light induces interactions between the molecules. The resulting hybrid light-matter states exhibit complex structure even if only a single vibrational coordinate per molecule is considered. Since all molecules are identical, it is appealing to start from the uniform situation where all molecules posses the same value of this vibrational coordinate. Then, polaritons and dark states follow like in atoms, but are functions of this coordinate, and this vibrational degree of freedom makes the physics different from that of atoms. However, in spite of all molecules being identical, each molecule does have its own vibrational coordinate. It is thus a vital issue to understand the meaning of the uniform situation and how to depart from it, and enable one to realistically investigate the ensemble. A rigorous and physically relevant meaning of the polariton energy curves in the uniform situation has been found. It is proven that any point on a  polariton energy curve is a (local) minimum or maximum for departing from the uniform situation. It is shown how to explicitly compute the energetic impact of departing from the uniform situation using solely properties of a single free molecule in the absence of the quantum light. The structure of the dark states and their behavior upon departing from the uniform situation is analyzed as well. Useful techniques not used in this topical domain are introduced and general results on, for example, minimum energy path, symmetry breaking and restoration, are obtained. It is shown how to transfer the findings to include several or even many nuclear degrees of freedom per molecule and thus to address the problem of quantum light interacting with many complex molecules. It is demonstrated that the interplay of several vibrational degrees of freedom in a single molecule of the
ensemble is expected to lead to additional and in part qualitatively different physics. General consequences are discussed.

\end{abstract}

%\vspace{1cm}
%\begin{tocentry}
%\begin{center}
%    \includegraphics[width=4.5cm]{schematic.eps}
%\end{center}
%\end{tocentry}

%{\Large \textbf{Graphical TOC Entry}}

%\vspace{0.5cm}

%\begin{center}
%    \includegraphics[width=5cm]{schematic.eps}
%\end{center}

\maketitle
%\newpage
\section{Introduction} \label{Introduction}
The coupling of matter excitations with quantized radiation field like that inside a cavity has become an extensive field of research. This coupling gives rise to the formation of hybrid light-matter states opening up new pathways to control and to manipulate static and dynamic properties of the matter. Ample work has been recently published in this field reporting on technical advances and motivated by the many possibilities to enhance or suppress available mechanisms and even to mediate new ones. Explicit examples are the possibilities to control photochemical reactivity \cite{Cavity_Chem_Reac_1,Cavity_Chem_Reac_2}, to control chemical reactions by varying the quantized field \cite{Cavity_Chem_Reac_3,Cavity_Chem_Reac_4,Cavity_Chem_Reac_5,Cavity_Chem_Reac_6}, to enhance energy transfer \cite{Cavity_CT_3,Cavity_ET_1} and charge transfer \cite{Cavity_CT_1,Cavity_CT_2,Cavity_CT_3,Cavity_CT_4}, to enhance or to completely suppress interatomic Coulombic decay \cite{Cavity_ICD}, and to induce new molecular non-adiabatic processes not available in free space  \cite{Cavity_Chem_Reac_5,Cavity_Non_Adiab_1,Cavity_Non_Adiab_2,Cavity_LICI_1,Cavity_LICI_3,Cavity_LICI_4,LICI_LiF_Huo,Cavity_Coll_CI}.  

In this work we investigate the molecular structure in the hybrid light-matter states formed by letting an ensemble of $N$ identical molecules interact with a quantized radiation field. We start with each molecule having a single vibrational degree of freedom as done before. In spite of this simplification, the situation turns out to be very complex. As each molecule has its own vibrational degree of freedom, and all the molecules of the ensemble interact with each other indirectly via the quantum radiation field, the situation resembles at first sight that of a macromolecule with $N$ coupled vibrations. The accurate treatment of a macromolecule is, in general, beyond reach. The situation is very different from that of an ensemble of molecules in a laser field, where each molecule with its single vibrational degree of freedom can be described separately. We develop a strategy to meaningfully study this complexity and bring some order into the findings by exploiting the symmetry of the ensemble and its wavefuctions. The focus here is on deriving rigorous results. We shall see that these results lead to new qualitative and quantitative insight into the hybrid light-matter states and their potential energy landscape. Then, we outline how to transfer the strategy developed for one vibrational degree of freedom per molecule to incorporate more or even all degrees of freedom of the molecules and thus open the possibility to investigate ensembles of truly polyatomic molecules interacting with quantum light. Thereby, it will become clear that the interplay of several vibrational and also rotational degrees of freedom in a single molecule of the ensemble can give rise to additional and in part qualitatively different physics.

\section{The Hamiltonian}
We consider an ensemble of $N$ non-interacting identical molecules in a cavity with a quantized light mode (cavity mode) of frequency $\omega_c$ and polarization direction $\vec{\epsilon}_c$. The total Hamiltonian of the ensemble-cavity system reads \cite{Cohen_Tannoudji_Book,Feist_PhysRevX,Oriol_Cavity_CP}:

\begin{align}\label{Ensemble-Cavity-Hamiltonian}
	H = H_e + \hbar\omega_c\hat{a}^\dagger \hat{a} + g_0 \vec{\epsilon}_c\cdot \vec{d}(\hat{a}^\dagger + \hat{a}),
	%\frac{-3 ({\bf {u}} \cdot {\bf {\hat{D}}}^N)({\bf {u}} \cdot {\bf {\hat{D}}}^M) +  {\bf {\hat{D}}}^N \cdot {\bf {\hat{D}}}^M }{R^3} + O(\frac{1}{R^4}),
\end{align}
where $H_e = \sum_{i=1}^{N} H_i$ is the electronic Hamiltonian of the ensemble, $\vec{d} = \sum_{i=1}^{N} \vec{d}_i$ is the total dipole operator of the ensemble and $g_0$ is the coupling strength between the cavity and the molecules. The quadratic dipole self-energy term is neglected as it is only of relevance for very strong coupling.  

Since we are dealing with molecules, the wavefunctions $\psi_{n}(R_i)$ of each non-interacting molecule depend on its nuclear coordinate $R_i$. As often done, we start by attributing a single coordinate to each molecule. Later on we move to the general case of several or all coordinates of the molecules. The frequency of the cavity mode is chosen to be resonant with the excitation energy between the ground $\psi_{0}$ and the first excited $\psi_{1}$ states of a molecule. The matrix representation of the Hamiltonian in the basis of noninteracting cavity-ensemble states and in the single-excitation space reads \cite{Cavity_Coll_CI}

\begin{align}\label{Matrix-Hamiltonian}
	\mathbf{\mathcal{H}}
	&=  \sum_{i=1}^{N} V_0(R_i)\mathbf{1} + \mathbf{H} \nonumber \\
	\mathbf{H} &= 
	\begin{pmatrix}
		\hbar\omega_c & \gamma(R_1) & \gamma(R_2) & \gamma(R_3) & \cdots \\
		\gamma(R_1) & \Delta(R_1) & 0 & 0 & \cdots\\
		\gamma(R_2) & 0 & \Delta(R_2)  & 0 & \cdots \\
		\gamma(R_3) & 0 & 0 & \Delta(R_3)  & \cdots \\
		     \vdots    &  \vdots   &   \vdots &   \vdots & \ddots   
	\end{pmatrix}
	,
\end{align}
where $\mathbf{1}$ is a unity matrix of dimension $N+1$, $V_0(R_i)$ is the ground state potential energy of the $i-th$ molecule, $\Delta(R_i) = V_1(R_i) - V_0(R_i)$ is the vertical change of potential energy upon excitation, and the dipole coupling of the i-th molecule to the cavity mode is $\gamma(R_i) = g_0d_{01}(R_i), d_{01}$ being the dipole transition moment of the molecule along the polarization axis of the cavity. 

As is obvious from the above expression for the Hamiltonian, the various molecules couple to each other indirectly via their direct coupling to the cavity. Before investigating further the properties of the solutions of this Hamiltonian, we wish to illuminate the impact of this indirect coupling, more precisely, we would like to transform the indirect coupling to become direct coupling. By block diagonalizing the matrix $\mathbf{H}$ one can decouple the cavity mode and obtain a matrix which describes the shifts and couplings among the molecules. In contrast to the diagonalization of a Hermitian matrix, there are infinitely many unitary
transformations which bring such a matrix into block diagonal form, in our case one block of dimension 1 which decouples from the $N\times N$ matrix block describing the molecules only. The conditions for making the block diagonal transformation unique are weak \cite{Block_Diagonalization}. If one has no special requirements concerning the transformation, it is natural to demand that the only action that it should perform is to bring $\mathbf{H}$ into block diagonal form. Interestingly, this very weak condition already defines uniquely the unitary transformation matrix and thus also $\mathbf{H}$. 

The block diagonalization can be done analytically in closed form for two molecules using directly the expressions in \cite{Block_Diagonalization}. The resulting block-diagonal $2\times2$ Hamiltonian describing the direct interaction between the molecules induced by the cavity reads
\begin{align}\label{Block-Matrix-Hamiltonian}
	\mathbf{\mathcal{H}_{BD}}
	&= 
	\begin{pmatrix}
		\Delta(R_1) + \gamma(R_1)x_1 & \gamma(R_1)x_2 + \gamma(R_2)x_1 \\
		\gamma(R_2)x_1 + \gamma(R_1)x_2 & \Delta(R_2) + \gamma(R_2)x_2  \\
		 \end{pmatrix} + \frac{\Delta(R_2) - \Delta(R_1)}{(1+q)^2}\mathbf{C} 	,
\end{align}
where the matrix $\mathbf{C}$ is given by 
\begin{align}\label{Block-Matrix-C}
\mathbf{C}
&= 
\begin{pmatrix}
	(x_1x_2)^2 & \frac{1}{2}x_1x_2(x_2^2 - x_1^2) \\
	\frac{1}{2}x_1x_2(x_2^2 - x_1^2) & -(x_1x_2)^2  \\
\end{pmatrix}.
\end{align}
Here, $x_i=\frac{\gamma(R_i)}{\lambda-\Delta(R_i)}$, $i=1,2$, and $q=(1+x_1^2 + x_2^2)^\frac{1}{2}$, where $\lambda$ is the eigenvalue of $\mathbf{\mathcal{H}}$ which emerges from cavity mode and becomes $\hbar\omega_c$ as the couplings $\gamma(R_i)$ approach zero (see Eq. (\ref{Matrix-Hamiltonian})). 

The structure of $\mathbf{\mathcal{H}_{BD}}$ shows that the dressing of the molecules by the field is rather involved due to the matrix $\mathbf{C}$. On the other hand, it is simple enough to be used for analysis and gaining insight into this dressing. Particularly illuminating and simple is the case where both molecules have the same value of the coordinate, $R_1=R_2\equiv R$. Then, $\Delta(R_2) - \Delta(R_1)$ vanishes and the matrix $\mathbf{C}$ does not contribute. The final result reads
\begin{align}\label{Block-Matrix-Hamiltonian-Simple-2}
	\mathbf{\mathcal{H}_{BD}}
	&= 
	\begin{pmatrix} \Delta(R) & 0 \\
		               0 & \Delta(R) \\ \end{pmatrix} + a\times\begin{pmatrix}
		1 & 1 \\
		1 & 1 \\
	\end{pmatrix}	,
\end{align}
where $a=\frac{1}{2}\left\{\frac{\hbar\omega_c-\Delta(R)}{2} - \left[2\gamma^2(R) + (\frac{\hbar\omega_c-\Delta(R)}{2})^2\right]^\frac{1}{2} \right\}$. We see that compared to two free molecules, at $R_1=R_2$, the cavity induces a direct coupling and shifts of the molecular energies which are identical. 

The latter result can be extended to $N$ molecules. The block matrix Hamiltonian takes on the appealing appearance
\begin{align}\label{Block-Matrix-Hamiltonian-Simple-N}
	\mathbf{\mathcal{H}_{BD}}
	&= 
	\Delta(R)\times\mathbf{1}_N + a_{{N}}\times\mathbf{11}_N , \nonumber \\	
	a_N &= \frac{1}{N}\left\{\frac{\hbar\omega_c-\Delta(R)}{2} - \left[N\gamma^2(R) + \left(\frac{\hbar\omega_c-\Delta(R)}{2}\right)^2\right]^\frac{1}{2} \right\}.
\end{align}
The quantity $\mathbf{11}_N$ is an $N\times N$ matrix where all its elements are $1$ and $\mathbf{1}_N$ is the unit matrix of the same dimension, see Eq.(\ref{Block-Matrix-Hamiltonian-Simple-2}). As for two molecules, also for $N$ molecules the cavity induces equal couplings and energy shifts for all molecules. 

\section{Uniformity and realistic example} \label{Uniformity and realistic example}
\subsection{General}
It is clear from the structure of the Hamiltonian matrix $\mathbf{H}$ given in Eq. (\ref{Matrix-Hamiltonian}) that all the $N$ molecules in the cavity contribute cooperatively to the resulting eigenstates. Since all the molecules are identical, it suggests itself to first investigate the uniform case where all molecules contribute identically. This is, of course, achieved at $R_1=R_2=...=R_N\equiv R$. We shall see later that this is a meaningful starting point to understand the complex potential energy surfaces in $N$ coordinates.   

To proceed, we notice that the Hamiltonian matrix $\mathbf{H}$ is an arrowhead matrix and its eigenvalues $E=\Delta(R) + e$ follow from the Dyson-like equation \cite{arrowhead_1,arrowhead_2,arrowhead_3}
\begin{align}\label{Dyson_1}
	e - \left[\hbar\omega_c - \Delta(R)\right] 
	&= 
	\sum_{i=1}^{N}\frac{\gamma^2(R_i)}{e - \left[\Delta(R_i) - \Delta(R)\right]}, 
\end{align}
where for convenience of later use, we have subtracted the constant $\Delta(R)$ for a reference $R$ from the diagonal of $\mathbf{H}$. 

In the uniform case, the above Dyson-like equation immediately reduces to $e - \left[\hbar\omega_c - \Delta(R)\right] = N\gamma(R)^2/e$. The resulting trivial quadratic equation leads to the two eigenvalues of the full Hamiltonian $\mathbf{\mathcal{H}}$ in Eq. (\ref{Matrix-Hamiltonian}):
\begin{align}\label{Polaritons_1}
	E_{u\pm} &= NV_0(R) + \frac{\hbar\omega_c+\Delta(R)}{2} \pm  \left[N\gamma^2(R) +\left(\frac{\hbar\omega_c-\Delta(R)}{2} \right)^2\right]^\frac{1}{2} .
\end{align}  
The respective states are known to be the upper $(+)$ and lower $(-)$ polaritonic states or just polaritons. Since we are dealing with molecules, we have added the subscript $u$ to indicate the uniform case. 

The general Dyson-like equation (\ref{Dyson_1}) possesses $N+1$ solutions. In the uniform case $N-1$ solutions decouple and simply give $e = 0$. This decoupling can also be seen directly in the effective Hamiltonian in Eq. (\ref{Block-Matrix-Hamiltonian-Simple-N}) operative in the molecular space only, where the matrix has rank 1 and $N-1$ eigenvalues of $11_N$ are zero. The resulting total energies of the decoupled states are
\begin{align}\label{Dark_1}
	E_{ud} &= NV_0(R) + \Delta(R).
\end{align}   
The respective $N-1$ degenerate states are called dark states as they do not couple to the cavity mode and cannot be reached from the molecular ground state by a dipole transition. We mention that the energies of the polaritons cannot cross those of the dark state manifold as a function of $R$. 

Polaritons and dark states have been amply discussed in the literature, in particular for atoms. For molecules the general solution of Eq. (\ref{Dyson_1}) or, equivalently, the eigenvalues of the Hamiltonian in Eq. (\ref{Matrix-Hamiltonian}) are rather complex functions in $N$ coordinate space and one cannot discern between dark and polaritonic states as one can do for atoms and the uniform case for molecules. All states can then be viewed as polaritons. Nevertheless, we will address states as dark or polaritons if it is clear from which kind of states in the uniform case they originate. The energies of the states in the uniform case are potential energy curves and in the general case they are potential energy surfaces. These curves and surfaces provide the potentials on which the nuclei move similar to the case of large free molecules. 

As for free molecules, where the minimum of a potential energy curve is a relevant quantity defining the equilibrium distance of the molecule in the state in question, we study the equilibrium distances $R_e$ of the polaritonic and dark states. First for the uniform case and later on we also attempt to relate these minima to those of the full potential energy surfaces in $N$ dimensional coordinate space. 

We begin the discussion with the dark states. For a single molecule in the cavity there is no dark state available. Dark states decouple from the cavity, but this does not imply that their energies are not affected compared to those of free molecules. Already for two molecules the single dark state has the energy $E_{ud} = 2V_0(R) + \Delta(R)$ which is nothing else than the sum of potential energies of the ground and excited molecular states $V_0(R) + V_1(R)$. The equilibrium distance of the dark state is thus between that of the ground and excited states of the free molecule. A glance at Eq. (\ref{Dark_1}) tells us that for many molecules in the cavity, the potential energy is dominated by that of the ground state potential, i.e., the equilibrium distance of the dark states is closer to that of the free ground state as the number of molecules in the ensemble grows. It is useful to introduce a minimal model, a so called linear-coupling model successfully used in many cases, see, e.g., \cite{Conical_Intersections_Review_1} and \cite{Cavity_Coll_CI}, where the potentials of the free molecules are harmonic and the coupling a linear function of $R$: $V_0(R) = \frac{1}{2}\omega R^2, \,\Delta(R) =  \Delta_0 + \kappa R , \, \gamma(R) = \gamma_0 + \beta R$. The equilibrium distance of the ground state is $R_{e0} = 0$ and that of the excited state $V_1 = V_0 +\Delta$ is $R_{e1} = -\kappa/\omega$. For that of the dark states one readily finds $R_{ed} = \frac{R_{e1}}{N+1}$.

To find the equilibrium distances of the upper and lower polaritons, one can resort to Eq. (\ref{Polaritons_1}) which provides the explicit expressions for the respective energies. For a general $N$, the resulting equilibrium distances follow only implicitly and depend more strongly on all the relevant parameters than for the dark states. For large $N$ an explicit expression can be obtained from
\begin{align}\label{Equilibrium_1}
	V_0^{'}(R) 
	&= 
	\mp\frac{\gamma^{'}(R)}{N^{1/2}} - \frac{\Delta^{'}(R)}{2N},
\end{align} 
where the primed quantities refer to derivatives with respect to $R$ and the upper sign refers to the upper polariton. In the minimal model, the equilibrium distances read: $R_{eu\pm} = \mp\frac{\beta}{\omega}\frac{1}{N^{1/2}} - \frac{\kappa}{\omega}\frac{1}{2N}$. In any case, it becomes clear that the quotient of the derivative of the coupling $\gamma$ and the ground state vibrational energy is the decisive quantity which determines the equilibrium distance for large $N$. Importantly, the minima of the polaritonic potentials approach that of the ground state potential of the free molecule as the number of molecules of the ensemble in the cavity grows. In this respect, small $N$ cases are more interesting as the results may vary with $N$ substantially.

\subsection{Realistic example}
\begin{figure}[h]
	\begin{center}
		\includegraphics[width=6cm]{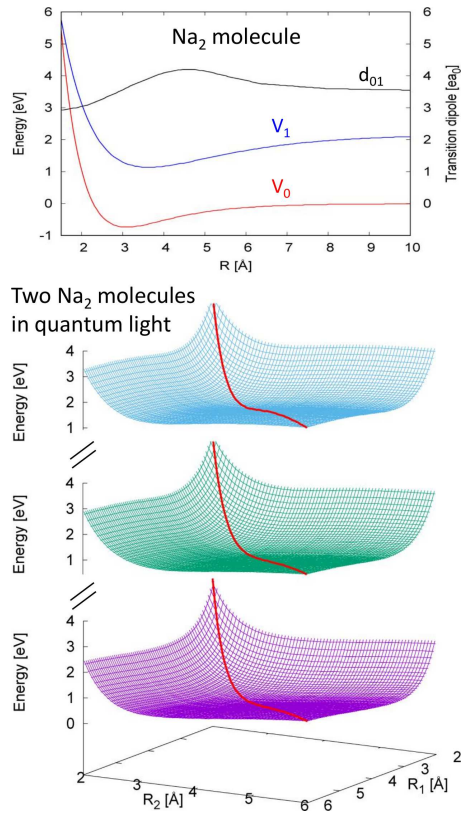}
		\caption{The potential energy surfaces of two sodium molecules in quantum light. Upper panel. The potential energy curves $V_0(R)$ and $V_1(R)$ of the ground and first excited electronic states of an isolated Na$_2$ molecule \cite{Sodium_Dimer_PECs} and the transition dipole moment $d_{01}$ between them \cite{Sodium_Dimer_Transition_Dipole}. These quantities serve as the input data for the calculation of the matrix Hamiltonian (\ref{Matrix-Hamiltonian}). Lower panel. Shown are the potential energy surfaces of the mixed matter-light states of two sodium molecules in a quantum field with coupling strength $g_0=3\times 10^{-3}$ au and frequency $\hbar\omega_c=2$ eV.  For each of the resulting three surfaces the respective uniformity energy curve is depicted in red.} 
		\label{fig:Two sodium molecules}
	\end{center}
\end{figure}

The sodium molecule Na$_2$ will serve as a show case example. A single sodium molecule has been investigated before in laser light \cite{LICI_1_2,LICI_4}, and in cavity \cite{Cavity_LICI_1}. Here, two and more sodium molecules will be considered. The potential energy curves of the ground X$^1\Sigma_u$ and first excited A$^1\Sigma_g$ electronic states as well as the transition dipole between these states as functions of internuclear distance are available \cite{Sodium_Dimer_PECs,Sodium_Dimer_Transition_Dipole} and depicted in the upper panel of Fig. \ref{fig:Two sodium molecules}. 

To obtain the potential energy surfaces of the polaritons and the dark state of two Na$_2$ molecules in quantum light, we employ the matrix Hamiltonian $\mathbf{\mathcal{H}}$ in Eq.(\ref{Matrix-Hamiltonian}) using the above data for the isolated molecule as input data. The eigenvalues of the resulting matrix have been computed for several values of the light-matter coupling $g_0$ and are shown in the lower panel of Fig. \ref{fig:Two sodium molecules} for a representative value of this coupling. 

The chosen frequency $\hbar\omega_c=2$ eV is very close to the energy difference $V_0(R)-V_1(R)$ at the equilibrium distance $R_{e0}$ of the ground electronic state of isolated Na$_2$ and thus, at first sight, all resulting three potential energy surfaces of the two sodium molecules look rather similar. In addition to the three energy surfaces, the respective uniformity energy curves are also depicted in the figure. Due to the choice of $\hbar\omega_c$, the energy split at the equilibrium distance between two successive surfaces and also between two successive uniformity curves is just given by the coupling $|\gamma(R_{e0})|$, see Eqs. (\ref{Polaritons_1},\ref{Dark_1}). Further discussion is given in the next section where also a rigorous meaning is provided for the uniformity energy curves. 

\section{Stability and symmetry breaking} \label{Stability and symmetry breaking}

The uniform case discussed above in Sec.\ref{Uniformity and realistic example} is physically appealing as all molecules are identical. However, as the resulting energies are one-dimensional curves embedded in $N$-dimensional coordinate space, the basic question arises whether we can give these energies a more precise meaning. The present section is devoted to answering this question. 

For the realistic example of two Na$_2$ molecules, the uniformity curves are seen to provide a minimum energy path, i.e., every point $R$ of the curves provides a local minimum in the perpendicular direction, see Fig. \ref{fig:Two sodium molecules}. These curves pass through the absolute minima of the full energy surfaces which constitute the equilibrium positions of the respective states. Is this always the case? 

\begin{figure}[h]
	\begin{center}
				\includegraphics[width=6cm]{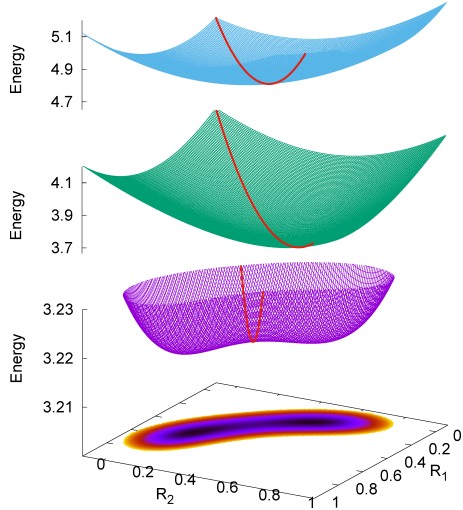}
		\caption{Illustrative example of two molecules in a quantum field demonstrating symmetry breaking. The figure shows the resulting three potential energy surfaces of the hybrid matter-light states where the matrix Hamiltonian (\ref{Matrix-Hamiltonian}) has been computed employing the minimal model of shifted harmonic oscillators explained in the text. The ground electronic state of the isolated molecule is $V_0=\omega R^2/2$, the shift is $V_1-V_0=\Delta=\Delta_0 + \kappa R$ and for simplicity the coupling $\gamma=\gamma_0$ is taken to be a constant. The values of the parameters in arbitrary energy units are: $\hbar\omega_c=\Delta_0=4$, $\kappa=-1$, $\gamma_0=0.5$  and the vibrational frequency $\omega=1$.  For each of the resulting three surfaces the respective uniformity energy curve is depicted in red.} 
		\label{fig:Symmetry_breaking}
	\end{center}
\end{figure}

In many problems studied the function under investigation is symmetric in multiple variables and at its minimum or maximum the variables are equal. A list of some examples is found in \cite{Purkiss_Principle}. Although this situation is likely to be present and can be proven for polynomial up to the third degree, it is not generally true. A simple example is provided by the fourth order symmetric polynomial $\left[(x-a)^2 + (y-b)^2\right]\left[(x-b)^2 + (y-a)^2\right]$ which for $a\neq b$ has two symmetry broken minima. Returning to the present problem of the eigen-energies of the Hamiltonian in Eq. (\ref{Matrix-Hamiltonian}), we have performed many numerical calculations until we found cases of symmetry breaking. An example for symmetry breaking employing this Hamiltonian within the minimal model introduced for analysis in Sec.\ref{Uniformity and realistic example} is depicted in Fig. \ref{fig:Symmetry_breaking}. In the figure it can be seen that the symmetry breaking occurs only for the lower polariton surface which exhibits two distinct minima at $R_1 \neq R_2$. We mention already here that the general conditions for symmetry breaking will be given explicitly in Sec.\ref{Dyson-like equation for stability and symmetry breaking}. 

To definitively answer the central question posed at the beginning of the section, we first quote a theorem \cite{Purkiss_Principle,Several_Variables}: \\
\underline{Theorem} (The Purkiss-Principle). Let $f$ and $g$ be symmetric functions with continuous second derivatives in the neighborhood of a point $P = (r,r,...,r)$. On a set where $g$ equals $g(P)$, the function $f$ will have a local maximum or minimum except in degenerate cases. 

We shall follow the idea of this theorem concretizing it for our problem of molecules in a cavity. The objects we are studying are the two energy eigenvalues $E_n$ of the Hamiltonian in Eq. (\ref{Matrix-Hamiltonian}) which originate from the polaritons. Due to their degeneracy, the situation is more complicated for the states originating from the dark states, except for the case of two molecules where there is only one dark state. The investigated energies are functions symmetric in the variables $R_1,R_2,...,R_N$. For $R_1=R_2=...=R_N\equiv R$ the respective energies are the uniformity curves in Eqs. (\ref{Polaritons_1},\ref{Dark_1}). For each point $P=(R,R,...,R)$ of an uniformity curve we investigate the behavior of the respective eigen-energy perpendicular to this curve in the vicinity of this point. To define this set of points, we note that a vector $\vec{P_R}=R(1,1,...1)$ can be assigned to the point $P$. For a vector $\vec{G}=(R_1-R,R_2-R,...,R_N-R)$ to be perpendicular to $\vec{P_R}$, one simply finds the condition $\sum_{i=1}^{N} (R_i-R) = 0$. To proceed, we define $g(R_1,...,R_N) = \sum_{i=1}^{N} R_i$ and note that $g(P) = NR$, i.e., the condition reads $g(R_1,...,R_N)=g(P)$.  

To find the extrema of the energy $E_n$ under the constraint of the above condition, we define the constrained energy function
  \begin{align}\label{Energy_Constrained}
  	\bar{E}_n
  	&= 
  	E_n + \lambda\sum_{i=1}^{N} (R_i-R),
  \end{align}
where $\lambda$ is a Lagrange multiplier. Now, we make use of the known fact that for a differentiable symmetric function all the derivatives are equal at the point $R_1=R_2=...=R_N=R$ \cite{Purkiss_Principle,Several_Variables}, i.e., at the point $P$ on the uniformity curve:  
\begin{align}\label{D(P)}
	\frac{\partial{E}_n}{\partial{R_1}}(P)
	&= 
	\frac{\partial{E}_n}{\partial{R_2}}(P)
	= ... 
	=	\frac{\partial{E}_n}{\partial{R_N}}(P),
\end{align}
and the gradient takes on the simple form $\nabla E_n(P)=a(R)(1,1,...,1)$ at the point $P$. It immediately follows that the gradient of the constrained energy vanishes, $\nabla \bar{E}_n(P)=0$ and the Lagrange multiplier becomes $\lambda=-a(R)$. 

Having found that any point on an uniformity curve provides an extremum of the energy along the coordinate space perpendicular to this curve at that point, we may now ask whether this extremum is a maximum or minimum or another kind of extremum, e.g., saddle point. As usual, we continue by expanding the energy around the point $P$ up to second order terms:
\begin{align}\label{Taylor_1}
	\bar{E}_n
	&= 
	E_n(P) + \vec{G}\mathbf{K}(P)\vec{G}^\dagger , \nonumber \\
    \mathbf{K}(P)	 
	&=	
	\bigg\{\frac{\partial^2{E}_n}{\partial{R_i}\partial{R_j}}(P)\bigg\}, \, \,  i,j=1,2,...,N,
\end{align}
where the row vector $\vec{G}$ has been defined above and has components $R_i-R$, and $\mathbf{K}(P)$ is the common Hessian matrix at the point $P=(R,R,...,R)$. 

To compute the Hessian we make again use of the known fact that for a twice differentiable symmetric function all the diagonal second derivatives are equal at the point P and this applies also to all the off-diagonal elements of the Hessian, i.e., 
\begin{align}\label{D2(P)}
\frac{\partial^2{E}_n}{\partial{R_i}\partial{R_i}}(P)
	&= 
	\frac{\partial^2{E}_n}{\partial{R_j}\partial{R_j}}(P) = c(R), \nonumber \\
	\frac{\partial^2{E}_n}{\partial{R_i}\partial{R_j}}(P)	 
	&=	
	b(R), \, \, i\neq j, \, \,  i,j=1,2,...,N.
\end{align}
These relations allow one to simplify the Hessian considerably. To proceed, we notice that the Hessian can straightforwardly be expressed as $\mathbf{K}(P)= [ \,c(R)-b(R)\, ]\times\mathbf{1}_N + b(R)\times\mathbf{11}_N$, where, as already introduced above, $\mathbf{11}_N$ is an $N\times N$ matrix where all its elements are 1 and $\mathbf{1}_N$ is the unit matrix of the same dimension. The similarity to the block-diagonal Hamiltonian in Eq.(\ref{Block-Matrix-Hamiltonian-Simple-N}) is striking. The Hessian is a matrix of rank 1 and has the eigenvalues $c(R)+(N-1)b(R)$ and $c(R)-b(R)$, the latter being $(N-1)$-fold degenerate. The respective eigenvectors can also be given explicitly. Assuming for simplicity an even number of molecules, they are $(1,1,...,1)/\sqrt{N}$ and $N-1$ vectors of the kind $(1,-1,...)/\sqrt{N}$ where the number of $+1$ and $-1$ elements is the same, respectively. 

To complete the derivation, we just transform the Hessian in Eq.(\ref{Taylor_1}) to its diagonal form using the above eigenvalues and eigenvectors. This transformation brings the coordinate vector $\vec{G}$ into a vector of collective coordinates $\vec{Q} = (Q_1,Q_2,...,Q_N)$ in which one obtains the final appealing result
\begin{align}\label{Taylor_2}
	\bar{E}_n
	&= 
	E_n(P) +  [c(R)-b(R)]\sum_{i=2}^{N}Q_i^2.  
	\end{align}
In these new coordinates the energy exhibits a simple quadratic form with a single prefactor which is either positive or negative definite (unless it vanishes accidentally) and as a consequence the extremum at P is either a minimum or a maximum. Saddle points and other appearances can be excluded. 

One should note that the coordinate $Q_1$ and the related eigenvalue  $c(R)+(N-1)b(R)$ of the Hessian do not appear in the final result as this coordinate is identically zero due to the constraint. The first two collective coordinates are
\begin{align}\label{Taylor_3}
	Q_1	 
	&=	
	(R_1+R_2+...+R_N-NR)/\sqrt{N}=0	,  \nonumber \\
	Q_2	 
	&=	
	(R_1-R_2+R_3-R_4+...)/\sqrt{N}, 
\end{align}
and the others follow from $Q_2$ by those permutations which result in linearly independent coordinates. 

The above findings make clear beyond doubt that the uniformity curves which have been introduced intuitively and which exhibit simple and explicit analytic expressions do posses a unique and important meaning as a relevant starting ansatz for investigating the energies of the ensemble of molecules in a quantum field in the $N$-dimensional coordinate space. 

Finally, let us briefly return to the heuristic example of Fig. \ref{fig:Symmetry_breaking}. For the states originating from the upper polariton and from the dark state, all the points on the respective uniformity curves are local minima and the global minimum of the exact energies is also lying on that curve. For the lower polariton the uniformity curve relates to minima at small and large values of $R$, but to maxima in between. Consequently, the exact energy exhibits symmetry breaking, where two equivalent absolute minima occur at $R_1 \neq R_2$. At the points along the uniformity curve where the minimum switches to a maximum, or vice versa, $c(R)=b(R)$ and the second-order term vanishes. The term 'except in degenerate cases' used in the theorem above, exactly applies to this situation.    

The next section is devoted to answer the question: is it possible to predict whether at a given point $R$ one encounters a local minimum or maximum solely based on the knowledge of the potential energies of the isolated molecule and its coupling to the cavity?

\section{Dyson-like equation for stability and symmetry breaking} \label{Dyson-like equation for stability and symmetry breaking}
\subsection{General and results on polaritons} \label{General and polaritons}
In this section we investigate the impact of distorting molecules from uniformity. Our major aim is to establish a quantitative expression for the curvature perpendicular to the uniformity curve and thus make direct connection to the basic finding in Eq. (\ref{Taylor_2}). Before doing so, we would like to mention that an analysis of the energy gradient arising due to distortion has been performed previously and has lead to the discovery of collective conical intersections of dark states  \cite{Cavity_Coll_CI}.

To investigate the impact of distortion of molecules from uniformity on the energies of the Hamiltonian, we start from the Dyson-like equation (\ref{Dyson_1}) which is equivalent to solving for the full Hamiltonian in Eq. (\ref{Matrix-Hamiltonian}). Having to fulfill the constraint $(R_1-R) + ... +(R_N-R)=0$ discussed in detail in the previous section, we have to distort at least two molecules. It should be stressed, however, that due to the simple quadratic structure found for the states originating from the polaritons, see Eq. (\ref{Taylor_2}), it is sufficient to distort only one pair of molecules in order to know whether one encounters a local maximum or a minimum of the energy at the value of $R$ in question. We start by distorting infinitesimally a pair of molecules fulfilling the constraint, say $R_1=R+\epsilon$ and $R_2=R-\epsilon$, and leaving the other $N-2$ molecules at $R$. As a consequence, the Dyson-like equation becomes
\begin{align}\label{Dyson-like-2}
	e - \left[\hbar\omega_c - \Delta(R)\right] 
	&= 
	\frac{\gamma^2(R+\epsilon)}{e - \left[\Delta(R+\epsilon) - \Delta(R)\right]} + \frac{\gamma^2(R-\epsilon)}{e - \left[\Delta(R-\epsilon) - \Delta(R)\right]} + \frac{(N-2)\gamma^2(R)}{e}. 
\end{align}

To continue, we expand the couplings $\gamma^2(R\pm\epsilon)$ and energy shifts $\Delta(R\pm\epsilon)$ in a Taylor series up to second order in $\epsilon$ and note that the Dyson-like equation above is equivalent to solving for the following auxiliary $4\times4$ matrix
\begin{align}\label{Auxiliary-Matrix}
	\mathbf{A}
	&=  \mathbf{B} + \mathbf{C} ,  \\
	\mathbf{B} &= 
	\begin{pmatrix}
		\hbar\omega_c - \Delta(R) & \gamma(R) & \gamma(R) & \sqrt{N-2}\gamma(R) \\
		\gamma(R) & 0 & 0 & 0  \\
		\gamma(R) & 0 & 0  & 0 \\
		\sqrt{N-2}\gamma(R) & 0 & 0 & 0  \\
		\end{pmatrix},  \nonumber  \\ 
	\mathbf{C} &= 
	\begin{pmatrix}
		0 & \gamma'\epsilon +\gamma''\epsilon^2/2 & -\gamma'\epsilon +\gamma''\epsilon^2/2 & 0 \\
		\gamma'\epsilon +\gamma''\epsilon^2/2 & \Delta'\epsilon +\Delta''\epsilon^2/2 & 0 & 0  \\
	-\gamma'\epsilon +\gamma''\epsilon^2/2	& 0 & -\Delta'\epsilon +\Delta''\epsilon^2/2  & 0 \\
       0 & 0 & 0 & 0  \\
	\end{pmatrix},\nonumber
\end{align}
where primed and double-primed quantities are their first and second derivative taken at $R$. As the matrix $\mathbf{B}$ is explicitly diagonalizable, the strategy to continue is to first diagonalize this matrix and then transform the auxiliary matrix $\mathbf{A}$ by the eigenvectors of $\mathbf{B}$. The result can then be used to calculate the impact of the distortions on the energies (and eigenstates). The calculation is lengthy, but straightforward. 

For completeness we list in the following the eigenvalues and respective eigenvectors of $\mathbf{B}$. Two eigenvalues are equal to $0$ and have the eigenvectors $(0,1/\sqrt{2},-1/\sqrt{2},0)$ and $(0,\sqrt{(N-2)/(2N)},\sqrt{(N-2)/(2N)},-2/\sqrt{2N})$. The other two eigenvalues are $X_\pm = (\hbar\omega_c-\Delta(R))/2 \pm  \left[N\gamma^2(R) +\left((\hbar\omega_c-\Delta(R))/2 \right)^2\right]^\frac{1}{2}$ and the respective eigenvectors read $(X_\pm y_\pm/\gamma, y_\pm,y_\pm,\sqrt{N-2}y_\pm)$, where $y_\pm = \gamma/\left[X^2_\pm +N\gamma^2(R)\right]^\frac{1}{2}$. The similarity between these eigenvalues and the energies of the polaritons and dark states in Eqs.(\ref{Polaritons_1},\ref{Dark_1}) is obvious. 

We remind that one has to add the potential energy $\sum_{i=1}^{N}V_0(R_i)+\Delta(R)$ to the solutions of the Dyson-like equation (i.e., the eigenvalues of the matrix $\mathbf{A}$ above) in order to obtain the eigen-energies of the full matrix Hamiltonian $\mathbf{\mathcal{H}}$ in Eq. (\ref{Matrix-Hamiltonian}). In the present case of distorting two molecules, this potential energy has also to be expanded as above giving $\sum_{i=1}^{N}V_0(R)+\Delta(R)+V''_0\epsilon^2$. The resulting matrix Hamiltonian describing the distortion of two molecules of the ensemble at any value of $R$ takes on the following appearance
\begin{align}\label{Distortion-Matrix-Hamiltonian}
	\mathbf{\mathcal{H}}
	 &= 
	 V''_0(R)\epsilon^2\mathbf{1} +
	\begin{pmatrix}
		E_{u+}+\alpha_+\epsilon^2 & 0 & \beta_+\epsilon & 0 \\
		0 & E_{u-}+\alpha_-\epsilon^2 & \beta_-\epsilon & 0  \\
		\beta_+\epsilon & \beta_-\epsilon & E_{ud}+\Delta''/2\epsilon^2  & \sqrt{\frac{N-2}{N}}\Delta'\epsilon \\
		0 & 0 & \sqrt{\frac{N-2}{N}}\Delta'\epsilon & E_{ud}+ \frac{N-2}{2N}\Delta''\epsilon^2  \\
	\end{pmatrix},  \nonumber  \\ 
	\end{align}
where the coefficients $\alpha_\pm$ and $\beta_\pm$ will be given below, and all the terms which give rise to changes of the eigenvalues up to and including second order in $\epsilon$ are retained. One sees that upon infinitesimal distortion in the space perpendicular to the uniformity curve, the upper and lower polaritons do not couple directly. Their coupling originates indirectly from their linear in $\epsilon$ coupling to the dark state. We shall return to the dark states below.

The total energies of the states of the Hamiltonian originating from the polaritons upon distortion away from the uniformity energies now read
\begin{align}\label{Polaritons_2}
	E_\pm 
	&= 
	E_{u\pm} + \{[V''_0(R) + \alpha_\pm] + [\beta_\pm^2 / (E_{u\pm} - E_{ud})]\} \epsilon^2 , \\
	\alpha_\pm
	&=
	[2\gamma''X_\pm/\gamma  + \Delta''] y^2_\pm \, \, , \, \, \beta_\pm = \sqrt{2}[\gamma'X_\pm/\gamma + \Delta'] y_\pm.  \nonumber
\end{align} 
All the quantities appearing in the above equation solely depend on the properties of the isolated molecule, i.e., $V_0, V_1$ and transition dipole, and the cavity, i.e., $g_0$ and cavity frequency $\omega_c$. Consequently, one can easily predict whether at a given value of $R$ one encounters a minimum or maximum of the total energies. The correction term upon distortion simplifies considerably for a resonance transition at $R$, when $\hbar\omega_c = \Delta(R)$. Then, assuming $\gamma > 0$ without loss of generality,  $X_\pm=\pm\sqrt{N}\gamma$ and $y^2_\pm = 1/(2N)$, and the term in front of $\epsilon^2$ takes on the explicit expression: $V''_0 \pm[\gamma''+(\gamma')^2)/\gamma]/N^{1/2} + [\Delta''/2 + 2\Delta'\gamma'/\gamma]/N \pm[(\Delta')^2/\gamma]/N^{3/2}$. As typically the curvature of the ground potential curve of molecules is positive $V''_0 > 0$, there is a tendency for the energies of the polaritons of large ensembles to exhibit minima in the relevant coordinate range, i.e., to avoid symmetry breaking. We shall return to this interesting behavior in the next subsection.

\subsection{Dark states} \label{Dark states}  
By inspecting the matrix Hamiltonian resulting upon distortion from uniformity, see Eq. (\ref{Distortion-Matrix-Hamiltonian}), one notices that for more than two molecules in the ensemble, in addition to the two polaritons, two dark states are affected by the distortion. Since there are $N-1$ dark states available, this implies that the remaining $N-3$ states of the ensemble stay dark upon the distortion of two molecules and have decoupled from the matrix. In contrast to the polaritons which are seen to change their energy by their joint interaction to a dark state, the appearing two states which originate from dark states are affected not only by the polaritons, but also by a direct interaction between them. This direct coupling leads to the lifting of their degeneracy found in the uniform case.  The situation is closely related to that discussed in \cite{Cavity_Coll_CI} where the lifting of the degeneracy of dark states gives rise to a collective conical intersection incorporating the coordinates of more than one molecule.   

Before further discussing the impact of distortion on the energies of the dark states for an ensemble of $N$ molecules, we first pay brief attention to the case of two molecules. For $N=2$ there is only a single dark state available. Indeed, one of the dark states is seen to decouple in the matrix Hamiltonian in Eq. (\ref{Distortion-Matrix-Hamiltonian}). The state originating from the dark state upon distortion follows immediately from the matrix and reads:
\begin{align}\label{Dark-state-2}
	E_d(N=2)
	&= 
	E_{ud} + \{[V''_0(R) + \Delta''/2] + [\beta_+^2 / (E_{ud}-E_{u+})] + [\beta_-^2 / (E_{ud}-E_{u-})]\} \epsilon^2 . 
\end{align} 
Clearly, the energy of the dark state is affected upon distortion by both polaritons. As the upper polariton pushes the dark state down and the lower polariton up in energy, there is some compensation of their joint impact. It should be noted that for N=2 the results of section (\ref{Stability and symmetry breaking}), in particular Eq. (\ref{Taylor_2}), are valid for the dark state as well as for the polaritons.

\begin{figure}[h]
	\begin{center}
				\includegraphics[width=6cm]{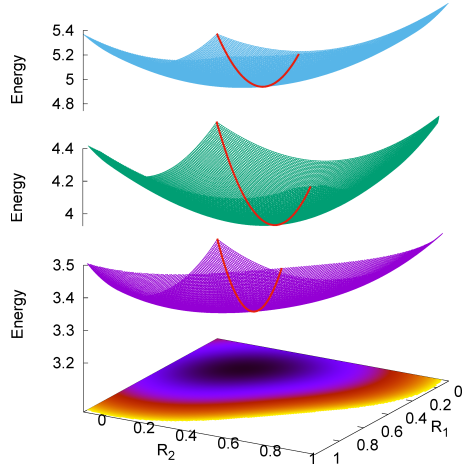}
		\caption{Illustrative example of two molecules in a quantum field demonstrating symmetry restoration by enlarging the vibrational frequency. The computation is as in Fig. \ref{fig:Symmetry_breaking} except that the vibrational frequency has been enlarged by a factor $1.5$. The symmetry breaking seen in Fig. \ref{fig:Symmetry_breaking} for the state originating from the lower polariton is restored. It should be noticed that according to Eq. (\ref{Two-molecules-simple}), decreasing $\kappa$ or enlarging the number of participating molecules can also lead to symmetry restoration. For each of the resulting three surfaces the respective uniformity energy curve is depicted in red.} 
		\label{fig:Symmetry_restoration}
	\end{center}
\end{figure}

In that respect, let us briefly comment on Fig. \ref{fig:Symmetry_breaking} which shows symmetry breaking occurring for the lower polariton. The situation is easily understood in the context of the minimal model of coupled shifted harmonic oscillators introduced in the paragraph above Eq. (\ref{Equilibrium_1}). To keep the comment short, we assume constant coupling around the value of $R$ investigated, i.e., $\gamma(R)=\gamma_0$, and choose the cavity mode to be resonant with the excitation at $R$, i.e., $\hbar\omega_c = \Delta(R)$. Then, $\beta_+ = \beta_-$ and as the energy gaps $E_{ud}-E_{u-}$ and $E_{ud}-E_{u+}$ are identical except of the sign, one finds the very simple result
\begin{align}\label{Two-molecules-simple}
	E_\pm(N)
	&= 
	E_{u\pm} + \bigg\{\omega \pm\frac{\kappa^2}{\gamma_0} \frac{1}{N^{3/2}}\bigg\} \epsilon^2, \nonumber \\
	E_d(N=2)
	&= 
	E_{ud} + \{\omega \} \epsilon^2. 	
\end{align}
It is seen that the states originating from the upper polariton and the dark state exhibit minima upon distortion, while this only applies to the one originating from the lower polariton if $\omega > \frac{\kappa^2}{\gamma_0} \frac{1}{N^{3/2}}$. Although $N=2$ for the dark state, we have kept the $N$ in the equation above for the polaritons to document that if there is symmetry breaking for any value of $N$, enlarging the number of molecules further will eventually lead to restoration of the symmetry of the lower polariton. An illustrative example is shown in Fig. \ref{fig:Symmetry_restoration} where the broken symmetry depicted in Fig. \ref{fig:Symmetry_breaking} is restored by choosing a larger vibrational frequency of the free molecule. 

To proceed with the evaluation of the two states originating from the dark states for general $N$, we return to the matrix Hamiltonian in Eq. (\ref{Distortion-Matrix-Hamiltonian}) and diagonalize the $2\times2$ lower block along the diagonal. The respective transformation changes the couplings $\beta_\pm$ to the polaritons. The resulting two states originating from the dark states split off from the remaining degenerate manifold of $N-3$ dark states. It is straightforward to arrive at the final expressions for the energies, but as these are lengthy we refrain from giving them explicitly. We concentrate instead on the large-$N$ case, where $(N-2)/N$ and $(N-2)/(2N)$ appearing in the matrix become essentially 1 and 1/2, respectively. The eigenvectors and eigenvalues of the $2\times2$ lower block of the matrix become trivial leading to the final result for the energies of the states originating from the dark ones: 
\begin{align}\label{Dark-states-large-N}
	E_{d1}
	&= 
	E_{ud} + \Delta'\epsilon + S \epsilon^2,  \\
	E_{d2}
	&= 
	E_{ud} - \Delta'\epsilon + S \epsilon^2, \nonumber \\
	S 
	&= 	[V''_0(R) + \Delta''/2] + [\beta_+^2 / (E_{ud}-E_{u+})]/2 + [\beta_-^2 / (E_{ud}-E_{u-})]/2.  \nonumber
\end{align}
Again, the quantity $S$ which determines the curvature of the energies upon distortion depends solely on the properties of a single molecule of the ensemble and of the cavity. In the minimal model of coupled shifted harmonic oscillators under the conditions discussed just above Eq. (\ref{Two-molecules-simple}), the energies simply become 
\begin{align}\label{Dark-states-large-N-simple}
	E_{d1,2} = E_{ud} \pm \kappa\epsilon + \omega \epsilon^2,
\end{align}
where $\omega$ is the vibrational frequency of the molecule. The lifting of degeneracy by the term $\pm \kappa\epsilon$ linear in the distortion coincides with that of the collective conical intersection found in \cite{Cavity_Coll_CI}, to which we refer for further characterization of this intersection in terms of Jahn-Teller interactions. Note that the energies of the states originating from the polaritons given in Eq. (\ref{Polaritons_2}) remain invariant by the above transformation.

Before closing this section, we discuss the impact of distorting more than one pair of molecules. From the general result in Eq. (\ref{Taylor_2}) it is clear that  we have found in the present section the answer to the question whether a point $R$ on the polaritonic curves is a local minimum or maximum. Nevertheless, it is illuminating to see explicitly the result of distorting more pairs. Let us start with four molecules and distort as above one pair by $R \rightarrow R \pm \epsilon_1$ and another pair by $R \rightarrow R \pm \epsilon_2$. The collective coordinates for distorting perpendicular to the uniformity curves are $Q_2=(R_1-R_2+R_3-R_4)/2, \,Q_3=(R_1+R_2-R_3-R_4)/2$ and $Q_4=(R_1-R_2-R_3+R_4)/2$. The result is $[c(R)-b(R)][\epsilon^2_1 + \epsilon^2_2]$ instead of $[c(R)-b(R)][\epsilon^2_1]$ when distorting a single pair. From this finding one can anticipate what happens if we distort $M$ pairs of molecules. To keep the calculation brief, we take the distortions of all $M$ pairs to be $R \rightarrow R \pm \epsilon$. 

The Dyson-like equation now takes on the appearance
\begin{align}
	e - \left[\hbar\omega_c - \Delta(R)\right] 
	&= 
	\frac{M\gamma^2(R+\epsilon)}{e - \left[\Delta(R+\epsilon) - \Delta(R)\right]} + \frac{M\gamma^2(R-\epsilon)}{e - \left[\Delta(R-\epsilon) - \Delta(R)\right]} + \frac{(N-2M)\gamma^2(R)}{e}, \nonumber
\end{align}
where we see that $M$ molecules are distorted from $R$ by $+\epsilon$ and $M$ by $-\epsilon$, and the remaining $N-2M$ molecules stay undistorted. Comparing with the Dyson-like equation (\ref{Dyson-like-2}) for a single distorted pair, one immediately sees that the new equation can be obtained from it by the following mapping $\gamma \rightarrow \sqrt{M}\gamma$ and $N \rightarrow N/M$. 

This mapping leaves $N\gamma^2$ invariant and also $X_\pm$, while $y_\pm \rightarrow \sqrt{M}y_\pm$, and thus $\alpha_+\pm \rightarrow M\alpha_+\pm$ and $\beta_+\pm \rightarrow M\beta_+\pm$. The ground states potential energy $\sum_{i=1}^{N}V_0(R_i)$ has also to be expanded as above for $M$ distorted pairs giving $\sum_{i=1}^{N}V_0(R)+MV''_0\epsilon^2$. Consequently, the final results for the total energies $E_\pm$ of the states of the Hamiltonian originating from the polaritons upon distortion away from the uniformity energies and those originating from the dark states $E_{d1,2}$ now read 
\begin{align}\label{Energies after distortion of M}
	E_{\pm}
	&= 
	E_{u\pm} +  M\times T \epsilon^2 , \\
	E_{d1,2}
	&= 
	E_{ud} \pm \Delta'\epsilon + M\times S \epsilon^2, \nonumber \\
	T
	&= [V''_0(R) + \alpha_\pm] + [\beta_\pm^2 / (E_{u\pm} - E_{ud})], \nonumber \\
	S 
	&= 	[V''_0(R) + \Delta''/2] + [\beta_+^2 / (E_{ud}-E_{u+})]/2 + [\beta_-^2 / (E_{ud}-E_{u-})]/2,  \nonumber
\end{align}
where for brevity we show the expressions for the dark states for $N>>2M$, see, Eq.(\ref{Dark-states-large-N}) and the discussion in the text above it. In other words, the changes of the energies quadratic in the distortion of $M$ molecular pair are $M$ times that for a single pair.

As a final remark on the states originating from the dark ones, we note that if all the molecules of the ensemble are distorted, i.e., $N=2M$, the coupling element in the matrix Hamiltonian (\ref{Distortion-Matrix-Hamiltonian}) between the dark states $\sqrt{(N-2)/N} \rightarrow  \sqrt{(N-2M)/N}$ vanishes. Therefore, one dark state decouples from all other states and another one originating from a dark state is found. This state, like we have discussed above in detail for $N=2$, is subject to the theorem of section (\ref{Stability and symmetry breaking}). The result for the energy of this state follows from that in Eq. (\ref{Dark-state-2}) and reads 
\begin{align}\label{Dark-state-for-all-pairs-distorted}
	E_d(N=2M)
	&= 
	E_{ud} + M\times \bar{S} \epsilon^2,  \\
	\bar{S} &= [V''_0(R) + \Delta''/2] + [\beta_+^2 / (E_{ud}-E_{u+})] + [\beta_-^2 / (E_{ud}-E_{u-})]. \nonumber 
\end{align}
For $N=2$, only a single dark state exists. If there are $N$ molecules in the ensemble and all are distorted in pairs, one dark state is quadratically affected by the distortion and the $N-2$ other dark states remain degenerate and are not affected. 

\section{Equilibrium bond lengths and wavefunctions} \label{Relations between equilibrium bond lengths and wavefunctions}
\subsection{On the symmetry of wavefunctions} \label{symmetry of wavefunctions}
In section \ref{Stability and symmetry breaking} we made ample use of the symmetry of the energies with respect to exchanging two molecules. In the present subsection we briefly discuss symmetry properties of the eigenvectors of the matrix Hamiltonian (\ref{Matrix-Hamiltonian}). In contrast to the Hamiltonian (\ref{Ensemble-Cavity-Hamiltonian}) which is symmetric and thus its eigenstates are either symmetric or antisymmetric, this does not apply to its matrix representation and the respective eigenvectors as the basis also depends on the coordinates of the molecules. As one can see from Eq. (\ref{Matrix-Hamiltonian}), exchanging two coordinates, e.g., $R_i \leftrightarrow R_j$, does not leave the matrix Hamiltonian invariant, but implies exchanging its $i$-th and $j$-th rows and columns. One can make use of this fact to learn more about the symmetry of the eigenvectors. 

Let us denote a normalized eigenvector of the matrix Hamiltonian by $\vec{\Psi}^\dagger = (\Phi_0,\Phi_1,...,\Phi_i, ..., \Phi_j,...\Phi_N)$, where, of course, its components are functions of the coordinates: $\Phi_k=\Phi_k(R_1,R_2,...,R_N)$. One can exchange $R_i$ and $R_j$ in two ways. The first is by exchanging the $i$-th and $j$-th rows and columns of the matrix Hamiltonian by a matrix transformation and the second is just by literally exchanging the two coordinates in the matrix. The transformation $\mathbf{T}$ which exchanges the $i$-th and $j$-th rows and columns of matrix Hamiltonian $\mathbf{\mathcal{H}}$, i.e.,$\mathbf{T}\mathbf{\mathcal{H}}\mathbf{T}$, is obtained from an unity matrix $\mathbf{1}$ by exchanging its $i$-th and $j$-th rows and columns: $T_{k,l}=\delta_{kl}, \, k,l\ne i,j$ and $T_{i,i}=T_{j,j}=0, \, T_{i,j}=T_{j,i}=1$. The transformed eigenvector takes on the structure $\vec{\Psi}^\dagger\mathbf{T} = (\Phi_0,\Phi_1,...,\Phi_j,...,\Phi_i,...\Phi_N)$, i.e., the $i$-th and $j$-th entries of the vector $\vec{\Psi}^\dagger$ are exchanged. 

Now, by exchanging the two coordinates $R_i \leftrightarrow R_j$, the transformed matrix Hamiltonian becomes the original matrix Hamiltonian and thus applying this exchange of coordinates also to the transformed eigenvector, recovers the original eigenvector. This immediately leads to the symmetry relations 
\begin{align}\label{Symmetry-Relations-Wavefunctions}
	\Phi_0(...,R_i,...,R_j,...)
	&= 
	\Phi_0(...,R_j,...,R_i,...), \\
	\Phi_i(...,R_i,...,R_j,...)
	&= 
	\Phi_j(...,R_j,...,R_i,...), \nonumber \\
	\Phi_k(...,R_i,...,R_j,...)
	&= \Phi_k(...,R_j,...,R_i,...), \, \, k\ne i,j. \nonumber 
	\end{align} 
The first entry of an eigenvector, $\Phi_0$, which relates to the cavity mode, is a symmetric function of the coordinates. The other components are symmetric functions as long as the exchanged entries $i,j$ differ from the assignment $k$ of this component. And, importantly, if a $k$-th component of an eigenvector equals one of the exchanged coordinate entries $i,j$, it becomes equal to the other component.  

As an interesting consequence of the above symmetry, an eigenvector of the matrix Hamiltonian can be expressed solely by two of its components, $\Phi_0$ and, for instance, $\Phi_1$, reading: $\vec{\Psi}^\dagger = (\Phi_0,\Phi_1,\Phi_1(R_1 \leftrightarrow R_2),\Phi_1(R_1 \leftrightarrow R_3),...,\Phi_1(R_1 \leftrightarrow R_N))$. 

\subsection{Forces and equilibrium} \label{Forces and equilibrium}
To calculate forces in atoms, molecules and crystals, the Hellmann-Feynman theorem \cite{Hellmann_Feynman_00,Hellmann_Feynman_01,Hellmann_Feynman_1,Hellmann_Feynman_2,Hellmann_Feynman_3} is amply employed. 
In the present subsection we apply this theorem to investigate the forces exerted on the nuclei of the molecules of the ensemble in quantum light. The way the Hellmann-Feynman theorem appears in the original works and in quantum mechanics textbooks is correct only for nondegenerate states. To keep the discussion compact, we focus here on the forces exerted when the molecules are in states originating from the polaritonic states. We mention, however, that the theorem has been explicitly extended for degenerate states in a form which can be also applied to states originating from dark states \cite{Hellmann_Feynman_4}.

To start with, we write the investigated total energy as an expectation value of the matrix Hamiltonian (\ref{Matrix-Hamiltonian}) taken with its respective eigenvector 
\begin{align}\label{Expectation-Value}
	E_n
	&= 
	\vec{\Psi}^\dagger \mathbf{\mathcal{H}} \vec{\Psi}, \\
	\vec{\Psi}^\dagger
	&= 
	(\Phi_0,\Phi_1,...,\Phi_N), \nonumber
	\end{align} 
and remind that each component of the normalized eigenvector is a function of all coordinates, $\Phi_k=\Phi_k(R_1,R_2,...,R_N)$ obeying the symmetry relations in Eq. (\ref{Symmetry-Relations-Wavefunctions}).

In the above matrix notation, the Hellmann-Feynman theorem states that the force exerted on the nuclei in the $i$-th molecule takes on the following appearance 
\begin{align}\label{Force}
	\frac{\partial{E}_n}{\partial{R_i}}
	&= 
	\vec{\Psi}^\dagger \frac{\partial{\mathbf{\mathcal{H}}}}{\partial{R_i}} \vec{\Psi}, 
\end{align} 
where the partial derivative of the matrix Hamiltonian implies that the derivative of each element of this matrix is taken.

As the partial derivative of the matrix Hamiltonian with respect to a single coordinate becomes a very sparse matrix, see Eq. (\ref{Matrix-Hamiltonian}),
the force has a simple appearance
\begin{align}\label{Force-Explicit}
	\frac{\partial{E}_n}{\partial{R_i}}
	&= 
	V'_o(R_i) + 2\Phi_0\Phi_i\gamma'(R_i) + \Phi^2_i \Delta'(R_i).
\end{align}
As already introduced above, the primed quantities denote the derivative of the respective quantity. All primed quantities depend on a single coordinate, and the cooperative interaction between the molecules enters through the eigenvector components $\Phi_0$ and $\Phi_i$ which in turn depend on all coordinates. The symmetry relations (\ref{Symmetry-Relations-Wavefunctions}) also determine the symmetries of the forces. For instance, for two molecules the force $\partial{E}_n/\partial{R_1}$ exerted on the first molecule at $(R_1,\, R_2)$ equals the force $\partial{E}_n/\partial{R_2}$ exerted on the second molecule at $(R_2,\, R_1)$

At equilibrium the force vanishes and one obtains an implicit relationship between the equilibrium geometry of the ensemble, i.e., the equilibrium coordinate $R^0_i$ of each molecule, and the respective molecular component ($\Phi_i$) of the wavefunction and the component ($\Phi_0$) related to the cavity mode at equilibrium. If there is no symmetry breaking, the solution has already been discussed in sections (\ref{Stability and symmetry breaking}) and (\ref{Dyson-like equation for stability and symmetry breaking}). Then, the solution is on the uniformity curves of the polaritons and there the wavefunctions or eigenvectors are known. The stationary points of the polaritonic curves themselves can also be obtained from the explicit expressions for the respective energies in Eq. (\ref{Polaritons_1}) and each point $R$ on these curves has been shown to be a minimum in the direction perpendicular to the curve. 

If there is symmetry breaking for a state originating from a polariton, the stationary points of the polaritonic curves themselves can, of course, also be obtained from Eq. (\ref{Polaritons_1}), and at some value of $R$ the respective uniformity curve has been shown to be a maximum in the direction perpendicular to the curve. In such cases, the implicit equations for equilibrium are particularly relevant as they, for example, tell us about the wavefunction components if we know the points of symmetry breaking and give us information about equivalent minima. As an explicit example we consider two molecules and assume symmetry breaking, i.e., $R^0_1 \ne R^0_2$. The two equations to fulfill at equilibrium are
\begin{align}\label{Symmetry-Breaking-2-Molecules}
	V'_o(R_i) + 2\Phi_0\Phi_i(R_1,R_2)\gamma'(R_i) + \Phi^2_i(R_1,R_2) \Delta'(R_i)
	&= 
	0, \, \, i=1,2\\
	V'_o(R_i) + 2\Phi_0\Phi_j(R_2,R_1)\gamma'(R_i) + \Phi^2_j(R_2,R_1) \Delta'(R_i)
	&= 
	0, \,\, i,j=1,2, \,\, i\ne j\nonumber
\end{align} 
where the second set of equations above follows from the first set by the symmetry relations (\ref{Symmetry-Relations-Wavefunctions}). If one finds a solution $(R_1=R^0_1, \,R_2=R^0_2)$ from the first set, it follows from the second set that also $(R_1=R^0_2, \,R_2=R^0_1)$ is a solution. This is, of course, not surprising, but shows how the symmetry relations determine how many equivalent minima are to be expected.   

Finally, we briefly discuss the case where the coupling element $\gamma$ varies very little in the vicinity of a stationary point. Then, it immediately follows from Eq. (\ref{Force-Explicit}) that at equilibrium one finds the explicit relations $\Phi^2_i = -V'_0(R^0_i)/\Delta'(R^0_i)$ between the equilibrium coordinates and the respective wavefunction components. Note that $\Phi^2_0$ follows directly from the fact that the eigenvector is normalized. In the minimal model of shifted harmonic potentials, these explicit relations simplify further giving explicitly $R^0_i = -\Phi^2_i \kappa / \omega$. 

\section{Orientation and polyatomic molecules} \label{Orientation and polyatomic molecules}
\subsection{Diatomic molecules and their orientation} \label{Diatomic molecules and their orientation}
The discussion of the previous sections concentrated on the situation where each molecule possesses a single internal coordinate representing a vibrational mode. As a free diatomic molecule has only one vibrational mode, one may be tempted to assume that the discussion has fully covered the issue of diatomic molecules in quantum light. This is, however, far from being the case. The coupling of a diatomic molecule to the cavity depends on the angle $\theta$ between the molecular axis and the polarization vector of the cavity. For molecules like Na$_2$ one has to substitute the coupling $\gamma(R_i)$ of the $i$-th molecule in the matrix Hamiltonian (\ref{Matrix-Hamiltonian}) by its orientation dependent coupling $\gamma(R_i) \rightarrow \gamma(R_i)\cos{\theta_i}$. 

One faces two situations. If the orientation of the molecules is fixed in space, the value of $\theta$ takes on a constant value and the discussion above remains as is in the case all molecules have the same orientation. The methodologies introduced in the previous sections can straightforwardly be extended to apply for different but fixed orientations. The situation changes drastically, if the molecules can rotate. Then, $\theta_i$ becomes a dynamic variable and to each molecule we have to assign two coordinates, $R_i$ and $\theta_i$ \cite{LICI_1,Cavity_LICI_1}. 

In the case of two coordinates per molecule, the situation complicates substantially. Already for a single molecule in quantum or classical light, the potential energies of the ground and the first excited electronic states exhibit a conical intersection in $R,\theta$ coordinate space and such intersections have been shown to lead to substantial non-adiabatic effects \cite{LICI_1,Cavity_LICI_1}. How to proceed ? In analogy to the case of one coordinate per molecule, we introduce here uniformity potential energy \emph{surfaces} where all molecules have the same value of the $R$ as well as of their orientation $\theta$. Then, the respective energy solutions of the matrix Hamiltonian are again two polaritons and $N-1$ dark states, but now the polaritons are functions of two variables:
\begin{align}\label{Polaritons_Diatomics}
E_{u\pm} &= NV_0(R) + \frac{\hbar\omega_c+\Delta(R)}{2} \pm  \left[N\gamma^2(R)\cos^2{\theta} +\left(\frac{\hbar\omega_c-\Delta(R)}{2} \right)^2\right]^\frac{1}{2}, \nonumber \\
 E_{ud} &= NV_0(R) + \Delta(R).
\end{align} 
Clearly, the dark states do not depend on the orientation. 

\begin{figure}[h]
	\begin{center}
				\includegraphics[width=10cm]{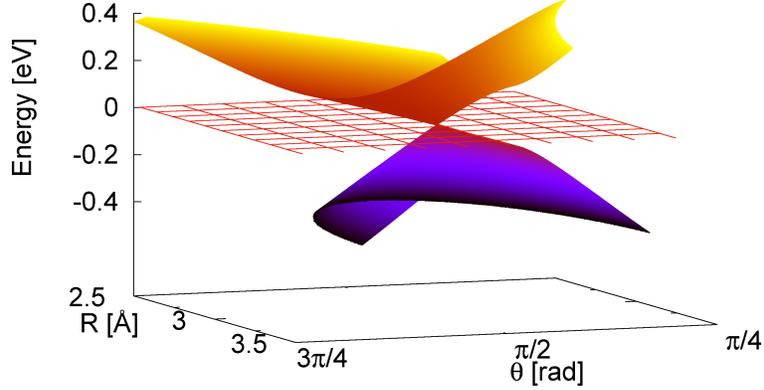}
		\caption{The uniformity potential energy surfaces of 10 sodium Na$_2$ molecules. For the sake of simplicity of the presentation, $9V_0(R) + V_1(R)$ has been subtracted from each surface.The molecules are allowed to rotate and thus, in addition to the vibrational coordinate $R$, also the angle $\theta$ between the molecular axis and the polarization direction of the quantum light are dynamical variables. The input data for the calculation is the same as employed in Fig. \ref{fig:Two sodium molecules}. All the 11 hybrid matter-light states are degenerate at $\theta=\pi/2$ and $R=R_c=R_{e0}$, the latter due to the choice of $\hbar\omega_c$. All 9 dark states are degenerate and independent of $\theta$ (Eq. (\ref{Polaritons_Diatomics})). The upper and lower polaritons exhibit a conical intersection.} 
		\label{fig:Uniformity_energy_surfaces}
	\end{center}
\end{figure}

The uniformity energy surfaces of an ensemble of $10$ Na$_2$ molecules are shown in Fig. \ref{fig:Uniformity_energy_surfaces} as a function of $R$ and $\theta$ for the same data as employed in Fig. \ref{fig:Two sodium molecules}. It is seen that as found for a single molecule \cite{LICI_1,Cavity_LICI_1}, the two polaritons exhibit a conical intersection at $\theta=\pi/2$ at the value of $R_c$ which fulfills $\Delta(R_c)=\hbar\omega_c$. Actually, all $N+1$ uniformity states intersect at this point. 

Searching for stationary points on the polaritonic surfaces along the coordinate $\theta$ gives the condition $\sin{2\theta}=0$ at values of $R \ne R_c$. At the solution $\theta=0$, the lower polariton takes on a minimum along $\theta$ and the upper polariton a maximum. The situation is just the opposite for the solution $\theta=\pi/2$. This coincides with the fact that the two surfaces repel each other along $\theta$ at any point $R \ne R_c$ and the repulsion is largest when the magnitude of the coupling $|\gamma(R)\cos{\theta}|$ is largest.

We anticipate that the role of the uniformity surface is similar to that we have discussed in detail for one coordinate per molecule, e.g., that it is a local maximum or minimum for distortions perpendicular to the surface, but as this work focuses more on one coordinate per molecule, we leave further discussions for the future. Nevertheless, before closing the subsection on diatomic molecules, we would like to apply the Hellmann-Feynman theorem to the orientation of the molecules of the ensemble.  In complete analogy to the result (\ref{Force-Explicit}), one finds that the force exerted on the orientation of a molecule is given by 
\begin{align}\label{Force-Explicit_Orientation}
	\frac{\partial{E}_n}{\partial{\theta_i}}
	&= 
	 2\Phi_0\Phi_i\gamma(R_i)\sin{\theta_i} .
\end{align}
At a stationary point, $\Phi_0 \ne 0$  for a state originating from a polariton, and hence either $\theta_i=0$ or $\Phi_i=0$. From the above analysis we conclude that  $\theta_i=0$, i.e., $\cos{\theta_i}=1$, corresponds to a maximum for a state originating from the upper polariton, and to a minimum if originating from the lower polariton. $\Phi_i= 0$ indicates that the $i-$th molecule does not contribute to the state in question. This, in turn happens when a molecule does not couple to the cavity mode, i.e., $\cos{\theta_i}=0$. As for free molecules, if we populate initially (e.g., by an external laser) an excited state, the forces will drive the system along a minimum energy path towards the minimum of the total potential energy. In cavity, one now has two relevant states instead of one, and it is important whether initially the state originating from the upper or lower polariton is populated. In the lower polariton, the force will drive the orientation to $\theta_i=0$, and if the state originates from the upper polariton, to $\theta_i=\pi/2$, i.e., the molecules will tend to reorient themselves such that they decouple from the cavity. 

\subsection{Strategy for polyatomic molecules} \label{Strategy for polyatomic molecues}
The results discussed in subsection (\ref{Diatomic molecules and their orientation}) on diatomic molecules throw light ahead on the large potential and palette of effects to be expected for realistic polyatomic molecules beyond the restricted dimensionality treatment. A general polyatomic molecule made of $N_{_N}$ nuclei possesses $3N_{_N}-6$ internal coordinates describing its vibrational modes and, in addition, $3$ Euler angles describing its orientation in space. How these Euler angles enter the matrix Hamiltonian is discussed in ref. (\cite{LICI_Polyatomic}) for classical light, and the outcome can be used for quantum light as well. Again, we can envisage two extreme cases where the molecule is fixed in space, for instance, by being embedded in some inactive material \cite{LICI_Polyatomic}, or it is free to rotate. In the former case, we have to treat only the internal coordinates and the specific fixed orientation gives rise to a specific coupling strength to the cavity mode as seen explicitly for diatomics above. In the latter case, the Euler angles become dynamical variables which one can add to the set of active coordinates. However, as the energies of the free molecule are independent of the orientation, one can calculate the forces exerted on the orientation of the molecules in the cavity explicitly as done above for diatomics.

In the following we restrict ourselves to molecules fixed in space. There are $N$ molecules in the ensemble and each has $M=3N_{_N}-6$ coordinates, and we introduce a supervector (i.e., a vector the elements of which are vectors) $\vec{\mathbf{\Re}}$ to describe the situation: $\vec{\mathbf{\Re}} = (\vec{R_1},...,\vec{R_N})$, where the $i$-th element $\vec{R_i}$ is a vector containing all the internal coordinates $R_{i\sigma}$ of the $i$-th molecule, $\sigma=1,2,...,M$.

In polyatomic molecules with their several or even many vibrational degrees of freedom, different electronic states are likely to come energetically close to each other somewhere in coordinate space. Indeed, even crossings of potential energy surfaces in the form of conical intersections are ubiquitous in polyatomic molecules \cite{Likelihood_CI,Conical_Intersections_Review_1}. At a crossing of $V_0$ and $V_1$ one cannot distinguish between the excited and ground state and the matrix Hamiltonian (\ref{Matrix-Hamiltonian}) has to be extended if the conical intersection is in the coordinate space of interest. We shall return to this interesting situation below and first discuss the situation where the ground and first excited electronic states do not exhibit a conical intersection and even do not come close to each other energetically. Formaldehyde (H$_2$CO) is an example of a molecule which has been recently investigated in cavity and does not exhibit a conical intersection between $V_0$ and $V_1$ \cite{Cavity_LICI_4}. Then, the matrix Hamiltonian in Eq. (\ref{Matrix-Hamiltonian}) is applicable if we substitute there the single coordinate $R_i$ of the $i$-th molecule by the vector $\vec{R_i}$ of the $M$ coordinates. 

As a first step towards characterizing and analyzing the $N\times M$ dimensional potential energy surfaces, we again introduce the uniformity energy surfaces where all molecules behave alike and $\vec{R_1}=...=\vec{R_N}=\vec{R}$. In the example of formaldehyde, this implies that the $C-O$ distance, the $H-H$ distance, \emph{etc.}, are identical in all the formaldehyde molecules of the ensemble. Of course, other internal coordinates than just interatomic distances can be used, even normal coordinates. The uniformity energy surfaces take on the explicit form
\begin{align}\label{Polaritons_Polyatomics}
	E_{u\pm} &= NV_0(\vec{R}) + \frac{\hbar\omega_c+\Delta(\vec{R})}{2} \pm  \left[N\gamma^2(\vec{R})+\left(\frac{\hbar\omega_c-\Delta(\vec{R})}{2} \right)^2\right]^\frac{1}{2}, \nonumber \\
	E_{ud} &= NV_0(\vec{R}) + \Delta(\vec{R}).
\end{align} 
While in the single coordinate per molecule case investigate in the previous sections a uniformity energy is a curve, i.e., one dimensional object, in the $N$-dimensional coordinate space, here, a uniformity energy surface is an $M$-dimensional hypersurface in the $N\times M$-dimensional space. It is left to the future to prove that any point on this hypersurface provides a local minimum or maximum of the energy upon distortion perpendicular to the surface like we could prove for the single coordinate per molecule case. If so, this will allow one to also derive the conditions for a minimum and for symmetry breaking based on the properties of the free molecule and cavity constant $g_0$ alone. As more than one coordinate per molecule are involved, light induced conical intersections can appear in polyatomic molecules even if the molecular orientation is fixed in space, see \cite{Cavity_LICI_4} for an example. For a single molecule, the equation above with $N=1$ provides the exact energy of the polaritons (there are no dark states). These polaritons exhibit a light induced conical intersection at the geometry which fulfills the conditions $\hbar\omega_c=\Delta(\vec{R})$ and vanishing coupling $\gamma(\vec{R})=0$. It is relevant to note that the polariton of the whole ensemble then exhibits a light induced conical intersection at the same location on the uniformity surface.  

In any case, the force exerted on the $i$-th molecule along one of its coordinates in an eigenstate of the Hamiltonian is just 
\begin{align}\label{Force-Explicit-Polyatomics}
	\frac{\partial{E}_n}{\partial{R_{i\sigma}}}
	&= 
	V'_o(R_{i\sigma}) + 2\Phi_0\Phi_i\gamma'(R_{i\sigma}) + \Phi^2_i \Delta'(R_{i\sigma}),
\end{align}
where $V'_o(R_{i\sigma})$ and the other primed quantities are partial derivatives with respect to the coordinate shown explicitly, but are functions of all the coordinates $\vec{R_i}$ of the $i$-th molecule and the eigenvector components $\Phi_0$ and $\Phi_i$ are functions of all the coordinates of the ensemble collected in $\vec{\mathbf{\Re}}$. 

It is interesting to extend the linear-coupling model of shifted harmonic oscillators to polyatomic molecules. Then, the ground state potential $V_0(\vec{R_i})$ of a molecule becomes a sum of harmonic oscillators of frequencies $\omega_{\sigma}$ and the energy shift $\Delta(\vec{R_i})=\Delta_0 + \vec{\kappa}\vec{{R_i}}$, where $\vec{\kappa}$ contains the linear couplings $\kappa_{\sigma}$ as elements, and analogously for the coupling $\gamma(\vec{R_i})$ of the molecule to the cavity mode. 
For simplicity, we consider here the minimal model where the $\gamma(\vec{R_i})$ is taken as a constant in the vicinity of equilibrium. At equilibrium the forces above vanish and one obtains from Eq. (\ref{Force-Explicit-Polyatomics}) for the equilibrium value of the coordinate $R^0_{i\sigma}=-\Phi^2_i\kappa_{\sigma}/\omega_{\sigma}$, or, equivalently $\Phi^2_i = |R^0_{i\sigma}||\omega_{\sigma}/\kappa_{\sigma}|$. Since the eigenstate vector is normalized $1 \geq \sum_{\i=1}^{N}\Phi^2_i$ and thus
$|\kappa_{\sigma}/\omega_{\sigma}| \geq \sum_{\i=1}^{N}|R^0_{i\sigma}|$, implying that for every vibrational mode, the sum of the \emph{absolute} values of all changes of the equilibrium coordinates of the whole ensemble from their ground state equilibrium value of the free molecules is restricted by the quotient $|\kappa_{\sigma}/\omega_{\sigma}|$ of that mode.  

What to do if there is a conical intersection of the potentials $V_0$ and $V_1$ in the coordinate space of interest? An example is pyrazine  \cite{Cavity_Pyrazine_Oriol_1,Cavity_Pyrazine_Oriol_2}. Then, clearly, the matrix Hamiltonian (\ref{Matrix-Hamiltonian}) is invalid and has to be extended to properly meet the situation. As usually done in conical intersection situations of polyatomic molecules, one resorts to a basis of so called diabatic electronic states \cite{Adiabatic_Diabatic_1,Adiabatic_Diabatic_2,Adiabatic_Diabatic_3} in which a matrix Hamiltonian is constructed \cite{Conical_Intersections_Review_1,Conical_Intersections_Book_1}. These electronic states change smoothly as a function of the vibrational coordinates, but the purely electronic matrix Hamiltonian is not diagonal anymore.

Around the conical intersection of $V_0$ and $V_1$ the respective states have to be taken into account on the same footing and this changes substantially the appearance of the matrix Hamiltonian. To visualize the situation, we consider here a single molecule and note that the same approach can be employed for an ensemble. For a single molecule the matrix Hamiltonian used until now consists of the $2 \times 2$ upper left corner of the matrix in Eq. (\ref{Matrix-Hamiltonian}). This matrix is the representation of the Hamiltonian operator (\ref{Ensemble-Cavity-Hamiltonian}) in the basis $\psi_01_c,\psi_10_c$, where $\psi_0$ and $\psi_1$ are the ground and excited electronic states of the molecule and $n_c, n=0,1$ denotes the number of cavity photons present. In the case of a conical intersection present already in the free molecule (so called natural conical intersection), we have to augment the above basis by $\psi_11_c,\psi_00_c$ in order to treat both states on the same footing. In addition, we now choose these electronic states to be the respective diabatic ones. The resulting matrix Hamiltonian is now $4 \times 4$ and reads
\begin{align}\label{Matrix-Hamiltonian-conical-one-molecule}
	\mathbf{\mathcal{H}}
	=
	\begin{pmatrix}
		\hbar\omega_c + V_1 & V_{01} & 0 & \gamma\\
		V_{01} & \hbar\omega_c + V_0 & \gamma & 0\\
		0 & \gamma & V_1 & V_{01}\\
		\gamma & 0 &  V_{01} & V_0
	\end{pmatrix}
	,
\end{align}
where $V_0$ and $V_1$ are the respective diabatic potentials and $V_{01}$ is the coupling potential \cite{Conical_Intersections_Review_1,Conical_Intersections_Book_1}, which, as well as $\gamma$, are functions of all the coordinates $R_\sigma, \, \sigma=1,2,...,M$ of the molecule.   

For a single molecule there are now four polaritons which exhibit various crossings and degeneracies. For instance, at the natural conical intersection, i.e., where $V_0= V_1$ and $V_{01}=0$, there are two doubly degenerate polaritons which split and interact with each other as the coordinates are varied. In general, there will be an interplay between the impact of the natural and of the light induced conical intersections on the dynamics of the molecule in the cavity.  

\section{Brief summary and outlook} \label{Brief summary and outlook}
The interaction of the molecules of an ensemble mediated by the quantum light renders the structure of the hybrid matter-light states rather complex even if a single vibrational degree of freedom is considered for each molecule. Having $N$ molecules implies the presence of $N+1$ hybrid matter-light states with $N$ coupled vibrational degrees of freedom in each of them. A rather hopeless situation to accurately attack without resorting to the underlying symmetry of the problem. The story simplifies considerably assuming uniformity where all molecules exhibit the same value $R$ of the respective vibrational coordinate. By doing so, the hybrid matter-light states and their energies can be given explicitly as functions of $R$ and can be viewed as polaritons and dark states at each value of $R$. But what is the meaning of these emerging polariton and dark energy curves? We prove a theorem which provides the rigorous physical meaning the uniformity curves get, showing what can be done in order to be able to compute the cooperative molecular structure in a straightforward way although the general issue is very complex. The uniformity curves turn out to have much physical content and lead to an ideal starting point for the analysis, understanding and more. In reality, each molecule has its own vibrational coordinate which can deviate from that of the others and lead to a departure from uniformity. The theorem tells us explicitly how the energy of the polaritons change upon departure from uniformity and provides us with knowledge of the potential energy landscape. In many cases, the polariton energy curves constitute a one-dimensional minimum energy path in the multidimensional coordinate space of the ensemble. Many results have been obtained like symmetry breaking and symmetry restoration and how they are explicitly determined by the properties of a single molecule in free space and of the cavity alone.     

The concept of uniformity can be carried over to account for more than one coordinate per molecule. In each degree of freedom, all molecules then posses the same value of the respective coordinate. This again leads to polaritons and dark states which can be given explicitly, but the respective energies are now functions of these values and constitute uniformity energy surfaces. Does the inclusion of additional degrees of freedom lead to qualitatively new physics already on this level of theory? The answer is clearly yes. As a first step we show that if diatomic molecules are free to rotate, already the uniformity surfaces in the vibrational-rotational coordinate space of the ensemble exhibit light-induced conical intersections (LICIs) stemming from that found for a single molecule \cite{Cavity_LICI_1}. Conical intersections give rise to non-adiabatic phenomena not present in a single degree of freedom (see also below) and such effects prevail when discussing the full problem beyond uniformity. Another interesting example is provided by the forces exerted by the quantum light on the orientation of the molecules. These forces are different in the lower and upper polaritons (and the dark states), leading to a decoupling of the molecules from the quantum light in the latter polariton. This behavior is found in both the uniformity and exact treatment.  

The uniformity surfaces also provide the starting point for analyzing and discussing the structure of an ensemble of general polyatomic molecules. They are multidimensional hyper-surfaces in the coordinate space of the ensemble. In contrast to diatomic molecules where the rotational degree of freedom is necessary to form the LICIs, such light-induced intersections are widely present in polyatomic molecules even at fixed orientation. It is shown that the LICIs present in a single polyatom molecule are transferred to the polaritons and that much more can happen when molecules depart from uniformity. We stress that LICIs and conical intersections in general have an enormous impact on the dynamics and spectroscopy of the molecules, qualitatively different from what is encountered in the absence of intersections \cite{Conical_Intersections_Review_1,Conical_Intersections_Book_1,Cavity_LICI_1,Cavity_LICI_2,Cavity_LICI_3,Cavity_LICI_4} and that they have a significant signature on the resulting level spacing statistics which reminds of classical chaos \cite{Level_Statistics_1,Level_Statistics_3,Level_Statistics_2}.

The results indicate an enormous wealth of possibilities and future potential for polyatomic molecules. Since conical intersections are omnipresent in free molecules, there is an interplay of the impact of these natural conical intersections and those induced by the quantum light. As conical intersections are a major reason for non-adiabatic processes, one can manipulate such processes by inducing LICIs in suitable positions in space and energy. Collective conical intersections in an ensemble have been predicted for the case of one degree of freedom per molecule once the ensemble contains at least 3 molecules \cite{Cavity_Coll_CI}. It is clear from the analysis in \cite{Cavity_Coll_CI} that this interesting finding prevails in polyatomic molecules with more than one active degree of freedom calling for the study of the resulting collective conical intersections. Other general future issues which deserve thorough investigations include nuclear dynamics, weak interactions between molecules and mixtures of molecules. The present investigation is on the molecular structure and provides access to the potential energy surfaces of the ensemble. It is of central importance to use the knowledge of the potentials in order to compute the classical and quantum dynamics on these potentials which are coupled by the LICIs and natural intersections in specific cases. To include the distortion of the molecules away from uniformity will be a challenging goal.  Usually, the molecules of the ensemble are assumed to be far away from each other so that they do not interact directly with each other. It will be relevant to include the weak long-range interactions between the molecules and to understand their impact on the potential energy surfaces of the hybrid matter-light states. Here, the symmetry of the wavefunctions can be exploited, but the situation is more involved than discussed in this work. Finally, it is rather obvious that on the long run there will be a need to investigate the impact of an impurity in the ensemble, an energy level of which participates in the formation of the hybrid matter-light states and, more generally, of a mixture of two (or more) kinds of molecules.

\begin{acknowledgments}
	The author thanks A. I. Kuleff and A. Vib\'ok for valuable contributions. Financial support by the European Research Council (ERC) (Advanced Investigator Grant No. 692657) is gratefully acknowledged
\end{acknowledgments}

\bibliography{biblio}

%\newpage

\end{document}